\documentstyle[prl,aps,epsfig,multicol]{revtex}

\begin{document}

\draft

\title{
Pion Interferometry of $\sqrt{s_{NN}} = 130$ GeV Au+Au Collisions at RHIC}
\author{
C.~Adler$^{11}$, Z.~Ahammed$^{23}$, C.~Allgower$^{12}$, J.~Amonett$^{14}$,
B.D.~Anderson$^{14}$, M.~Anderson$^5$, G.S.~Averichev$^{9}$, 
J.~Balewski$^{12}$, O.~Barannikova$^{9,23}$, L.S.~Barnby$^{14}$
% ...
\\
(STAR Collaboration)
}
\address{
% ...
$^5$University of California, Davis, California 95616       \\
% ...
$^{9}$Laboratory for High Energy (JINR), Dubna, Russia       \\
% ...
$^{11}$University of Frankfurt, Frankfurt, Germany       \\
$^{12}$Indiana University, Bloomington, Indiana 47408       \\
% ...
$^{14}$Kent State University, Kent, Ohio 44242       \\
% ...
$^{23}$Purdue University, West Lafayette, Indiana 47907       \\
% ...
}

\date{\today}
\maketitle

\begin{abstract}
Two-pion correlation functions in Au+Au collisions
at $\sqrt{s_{NN}} = 130$~GeV have been measured by the
STAR (Solenoidal Tracker at RHIC) detector.
The source size extracted by fitting the correlations
grows with event multiplicity and decreases with transverse momentum.
Anomalously large sizes or emission durations, which have been suggested as signals of quark-gluon plasma
formation and rehadronization, are not observed.
The HBT parameters display a weak energy dependence over a broad range in $\sqrt{s_{NN}}$.
\end{abstract}

\begin{multicols}{2}
\narrowtext

Two-particle intensity interferometry techniques (HBT) have been used
extensively to probe the space-time structure of heavy ion collisions~\cite{hbt-review}.
At midrapidity ($y = 0$) and low transverse momentum ($p_T$), two-pion correlation functions
reflect the space-time geometry of the emitting source, while dynamical information (e.g.
collective flow) is contained in the momentum dependence of the apparent source
size~\cite{hbt-flow,MS88,pratt-flow-and-lifetime}.

In this Letter we study two-pion correlation functions in
Au+Au collisions at
nucleon-nucleon center-of-mass energy 
$\sqrt{s_{NN}} = 130$~GeV
produced by the RHIC facility at Brookhaven National Laboratory,
and
compare our results to similar studies 
at lower energy.  Such systematic
comparisons may reveal the onset of new phenomena, as 
$\sqrt{s_{NN}}$
increases.
We are particularly interested in possible indications that a quark
gluon plasma (QGP) has been formed in the collision.
Several authors~\cite{hbt-review,pratt-flow-and-lifetime,rischke-lifetime,rischke-gyulassy-lifetime,bertsch89,shuryak-lifetime}
have proposed HBT studies to probe for a significant increase in the pion emission timescale associated with QGP formation.

Experimentally, the two-particle correlation function is obtained from
the ratio $C_2({\bf q}) = A({\bf q})/B({\bf q})$ (normalized to unity at large ${\bf q}$), where $A({\bf q})$ 
is the measured two-pion 
distribution of pair momentum difference ${\bf q=p_2-p_1}$, and $B({\bf q})$
is the mixed background distribution~\cite{event-mixing}, calculated in the same way
using pairs of particles taken from different events.

The STAR time projection chamber (TPC)~\cite{star-expt,tpc} was used to record charged
particle tracks as they crossed up to 45 padrows of the detector.
For P-10 gas in the 0.25~T solenoidal
magnetic field, transverse and longitudinal diffusion were both about
350$\mu {\rm m}$ times the square root of the drift length.
The TPC was read out with 138,000 waveform
digitizer channels~\cite{fee}; the signal shaping width was comparable
to the diffusion.  For a typical 1.5-meter drift, two-track resolution
was limited by the resulting 6 mm signal spread, in addition to 
geometric spreading due to tracks traversing the padrows at finite crossing angles.
This
analysis used tracks  with $0.125 < p_T < 0.45$~GeV/c and $|y|<0.5$,
for which the reconstruction and $\pi$ identification efficiencies are high.
We present results for three $p_T$ ranges: $0.125-0.225$~GeV/c, $0.225-0.325$~GeV/c, and $0.325-0.45$~GeV/c.

We select events 
with a collision position that is within 75~cm of the mid-plane of the 4-m long TPC.
An important feature of two-particle correlation functions is
that single-particle phase space and acceptance effects 
cancel to first order.
To preserve this feature in the collider environment (in which the collision vertex-- and hence the single-particle
acceptance-- varies event-to-event), we only mix events which have a primary
vertex longitudinal location within 15~cm of each other; however,
mixing events over the full $\pm 75$~cm range introduces 
no significant distortion in the correlation fit parameters, in the present analysis.

Starting with a minimum-bias trigger~\cite{star-elliptic},
we characterize the centrality of collisions off-line according to the measured multiplicity of
negatively-charged particles with pseudorapidity $|\eta|<0.5$.  After accounting for event
reconstruction inefficiency for low-multiplicity events, we estimate that the minimum-bias
distribution contains $\sim$90\% of the hadronic Au+Au cross section~\cite{star-elliptic}.
We analyze $\sim10^5$ events in each of three bins;
bin 3 contains the 12\% most central of the measured collisions, bin 2 the next 20\%, and bin 1 the next 40\%.

    Particle identification was achieved by correlating the magnetic rigidity of a track with its
specific ionization ($dE/dx$) in the 
gas of the TPC.
In the momentum region of interest, $\pi$, $K$, and $p/\overline p$ are well-separated; contamination
of the pion sample by electrons is the primary concern.  
Based on simulations and on extrapolation from regions of clear $e^-/\pi^-$ separation,
we estimate that electrons comprise about 10\% (4\%) of the selected tracks in our lowest (highest) $p_T$ bin.
Tracks used in the correlations are required to project back to the primary interaction vertex within 3~cm, thereby
selecting primarily pions emitted directly from the collision.
Contamination from non-pions and pions
originating from
long-lived decays (e.g. $\Lambda$, $\omega$)
reduces the strength of the correlation, characterized by the parameter $\lambda$,
while leaving the extracted source radii unchanged~\cite{E895HBT,Appelhauser}.

    Although track-splitting
(incorrect reconstruction of a single
particle as two particles) 
 in the STAR TPC is a small effect in general 
($\lesssim 1\%$),
it may have a strong impact on
correlation studies at low 
$|q|$.
False pairs generated by track-splitting
are removed by a topological cut 
which requires valid and distinct
signals of both tracks on several TPC padrows.
For the present analysis, residual effects of track-splitting
are negligible.

   To eliminate the effect of track-merging (in which two tracks with similar momenta are reconstructed as a single track),
 we require the tracks in a pair to be well
separated ($> 2.5$~cm at the radius of the TPC inner field cage, 50~cm).  In applying this cut also to
the mixed-pair background, the effect of event-to-event variation in primary vertex position is 
taken into account.
The anti-merging cut removes significant detector effects, but also discriminates against low-{\bf $|q|$} pairs,
which carry the correlation signal.
This leads to an artificial reduction in the HBT parameters, which we estimate, based on
simulations~\cite{merging-note}, to be 3-6\% for the radii and 6-14\% for $\lambda$,
depending on $p_T$.
We correct for this effect in the results presented.

We apply 
to each background pair
a Coulomb correction~\cite{pratt-flow-and-lifetime}
corresponding to
a spherical Gaussian source of  5~fm  radius; this correction is identical to that used
by several other groups~\cite{E895HBT,E877HBT,NA49Coulomb}.
In principle, this procedure over-corrects the correlation function
for realistic sources (e.g. in which some pions
originate far from the core source). 
We find that the HBT radii change smoothly by $\sim10\%$ as the strength of the Coulomb correction
is varied from the standard correction to no correction.
Here, we restrict ourselves to
the standard Coulomb correction,
allowing for
a uniform extension of existing interferometry systematics from lower energy.

One-dimensional correlation functions
constructed in 
the invariant quantity 
$Q_{inv}=\sqrt{({\bf p}_1-{\bf p}_2)^2-(E_1-E_2)^2}$
are fit to the functional form
\begin{equation}
\label{1dfit}
C(Q_{inv}) = 1 + \lambda {\rm exp}(-Q_{inv}^2R_{inv}^2)  .
\end{equation}
The 1-D correlation function (and fit) corresponding to our highest multiplicity class
is shown in Fig.~1a for low-$p_T$ $\pi^-$.
The 1-dimensional fit fails in
lowest $Q_{inv}$ bins, as has been observed in previous measurements~\cite{NA44HBT,WA98HBT}.
Hence, the extracted radius
$R_{inv}\approx 6.3$~fm  is only a rough indication of the space-time extent of the source.

    A more detailed characterization of the emitting source is obtained through multidimensional
correlation functions.  We decompose the momentum difference ${\bf q}$ according to the
Pratt-Bertsch~\cite{pratt-flow-and-lifetime,bertsch89} ``out-side-long'' (indicated by
$o$, $s$, and $l$ subscripts)
parameterization.  Here, $q_{l}$
is parallel to the beam,
$q_{s}$ is perpendicular to the beam and to the total momentum of the pair,
and $q_{o}$ is perpendicular to  $q_{l}$ and  $q_{s}$.
Data are analyzed in the longitudinally co-moving system (LCMS) frame, in which the longitudinal
component of the pair momentum vanishes.
Figs.~1b-d show one-dimensional
projections of the correlation function $C(q_o,q_s,q_l)$ onto the $q_o$, $q_s$, and $q_l$ axes,
for $\pi^-$ from the
most central collisions.

\begin{figure}[t]
\vspace*{-0.9cm}
\begin{center}
\epsfig{file=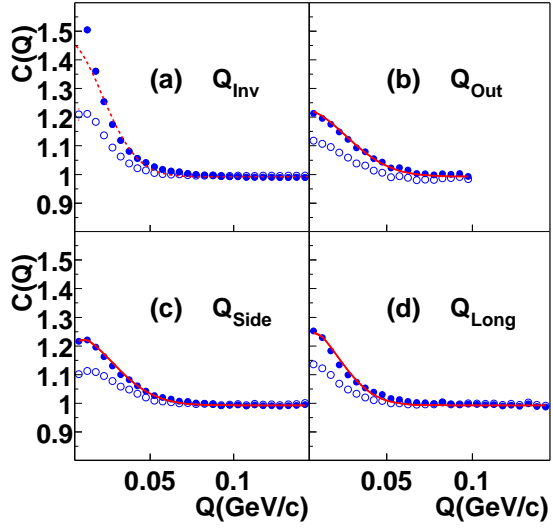,width=8cm}
\end{center}
\vspace*{-0.2cm}
\caption{Coulomb-corrected (full dots) and uncorrected (circles) correlation
functions for low-$p_T$ $\pi^-$ emitted at midrapidity from central collisions.
Shown in panel (a) is the $Q_{inv}$ correlation function, and in panels (b-d) 
projections of the 3-dimensional
correlation function onto the $q_{o}$, $q_{s}$, and $q_{l}$ axes.  
To project onto one $q$-component, the others are integrated over the range
0-35~MeV/c.
Fits to Coulomb-corrected correlations are shown by curves.}
\label{fig:CorrFctns}
\end{figure}

    The 3-dimensional correlation functions are fit with the standard Gaussian form
\begin{equation}
\label{eq:bertsch-pratt}
C(q_{o},q_{s},q_{l})=1+
\lambda {\rm exp}(-R_{o}^2q_{o}^2-R_{s}^2q_{s}^2-R_{l}^2q_{l}^2) ,
\end{equation}
where $R_i$ is the homogeneity length in the $i$~direction~\cite{hbt-review}.
Projections of the fit to the central collisions, weighted according to the mixed-pair background,
are shown as curves in Fig.~\ref{fig:CorrFctns}.

Systematic errors on the HBT fit parameters
have two sources of roughly equal magnitude:
(1) the uncertainty on the correction for the anti-merging cut, estimated equal to the correction itself, and
(2) the uncertainty associated with the Coulomb correction, $\sim$0.1-0.2~fm in the radii, determined by
varying the Coulomb radius by $\pm$1~fm.

    The effect of the single-particle momentum resolution ($\delta p/p\sim$~2\% in the TPC for the particles
under study) is known to induce systematic underestimates of HBT parameters.  
For our lowest $p_T$ bin, $\delta q_{o} \approx 4.5$~MeV/c, and $\delta q_s \approx \delta q_l \approx 3$~MeV/c. 
While $\delta q_s$ and $\delta q_l$ change little, in our highest $p_T$ bin, $\delta q_o \approx 11$~MeV/c.
Using an iterative procedure similar to that used at lower energies~\cite{E895HBT,NA44HBT},
we have corrected our correlation functions for finite resolution effects.  The correction
had no effect on the HBT parameters for our lowest $p_T$ cut; for the highest $p_T$ cut,
it resulted in a 5\% increase in $\lambda$ and a 8\% increase in $R_o$, while $R_s$ and $R_l$ were
unaffected.

\begin{figure}
\vspace*{-0.9cm}
\begin{center}
\epsfig{file=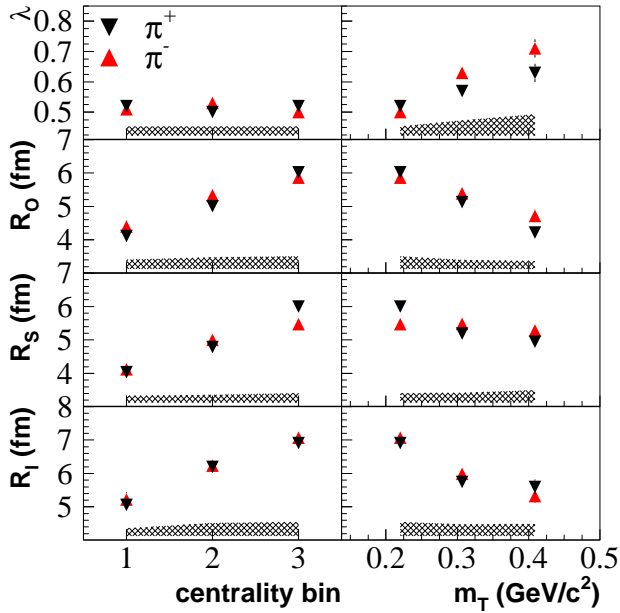,width=9cm}
\end{center}
\vspace*{-0.2cm}
\caption{
The multiplicity dependence of the HBT parameters
is shown for low-$p_T$ pions on the left; bin~3 contains
the high-multiplicity events.
For high-multiplicity collisions, the $m_T$-dependences are shown on the right.
Error bars indicate statistical uncertainties;
systematic uncertainties are
represented by the height of the shaded regions below.
}
\label{fig:Centrality-pTDep}
\end{figure}

    The multiplicity dependence of the source parameters
is presented in the left panels of 
Fig.~\ref{fig:Centrality-pTDep},
for low-$p_T$, midrapidity $\pi^-\pi^-$ and $\pi^+\pi^+$ correlations.
The parameters for positive and negative pions are
similar; the $\lambda$ parameter is constant, and
all three radii increase monotonically with multiplicity.
The increase of the transverse radii $R_{o}$ and $R_{s}$ with multiplicity
is interpreted as a geometrical effect, and is also observed in lower
energy measurements~\cite{E895HBT,E866HBT,NA49HBT-QM99}.
The multiplicity dependence of $R_l$ differs from observations at lower energies;
E895~\cite{E895HBT} and NA49~\cite{NA49HBT-QM99} observe no dependence of $R_l$ over a wide range of multiplicity
at the AGS and SPS, while NA44~\cite{NA44-CentralityHBT} reports a sharp increase for the very highest
multiplicity collisions at the SPS.

    The right panels of Fig.~\ref{fig:Centrality-pTDep} show, for the events in multiplicity bin 3, the 
dependence of the HBT parameters on $m_T = \sqrt{p_T^2+m^2}$.  $\lambda$ increases with $m_T$,
consistent with studies at lower energy~\cite{E895HBT,NA44HBT,WA98HBT},
in which the increase was attributed to decreased contributions of pions from long-lived resonances at higher $p_T$.

The radius parameters decrease significantly with $m_T$.
This $m_T$-dependence, weak at $\sqrt{s_{NN}}\sim$~2~GeV~\cite{E895HBT} and growing
stronger with collision energy~\cite{NA44HBT,WA98HBT,E866HBT,NA49HBT-QM01},
reflects pion emission from a radially expanding source~\cite{hbt-review}.  The
$m_T$-dependence at RHIC is similar, but not identical, to that observed in central Pb+Pb collisions at the 
CERN SPS~\cite{NA44HBT,WA98HBT,NA49HBT-QM01}.  
The significant decrease of $R_o$ with $m_T$ contrasts with the
   $m_T$-independence of $R_o$ measured at midrapidity in 
   collisions at SPS energy~\cite{WA98HBT,NA49HBT-QM01}.
For our highest $m_T$ data, $R_o < R_s$, possibly indicating significant source opacity resulting from the high
particle densities generated in RHIC collisions~\cite{hv98,th98}.

\begin{figure}
\vspace*{-0.9cm}
\begin{center}
\epsfig{file=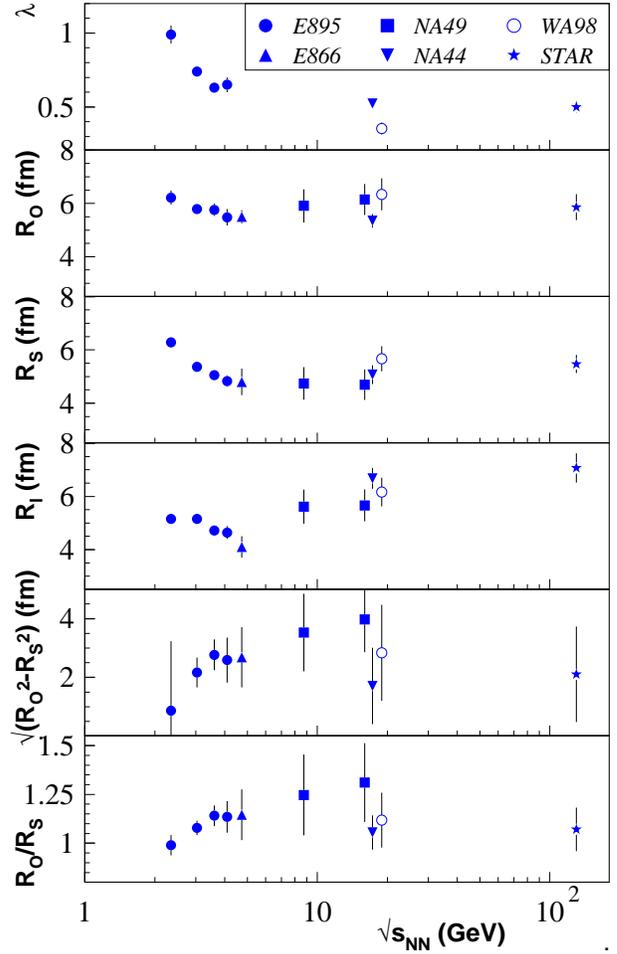,width=9cm}
\end{center}
\vspace*{-0.2cm}
\caption{
The energy dependence of $\pi^-$ HBT parameters for central Au+Au (Pb+Pb) collisions at
midrapidity and $p_T\approx$~0.17~GeV/c~[\ref{lab-E895},\ref{lab-NA44},\ref{lab-WA98},\ref{lab-E866},\ref{lab-NA49QM01}].
The SPS data are offset slightly in $\sqrt{s_{NN}}$ for clarity.
Error bars on NA44, NA49, and STAR results include systematic uncertainties; error bars
on other results are statistical.
}
\label{fig:ExcitationFctn}
\end{figure}

To place our results in context, Fig.~\ref{fig:ExcitationFctn} shows the world's dataset
of correlation parameters for midrapidity, low-$p_T$ ($\sim 170$~MeV/c) $\pi^-$ from central
Au+Au or Pb+Pb collisions.
The parameter $\lambda$ falls smoothly and rapidly from unity (the expected value
from na\"ive
assumptions) at $\sqrt{s_{NN}}\sim$2~GeV, to about 0.5 at RHIC; this decrease
is attributed partially to an increased fraction of $\pi^-$
arising from long-lived 
decays as $\sqrt{s_{NN}}$ increases~\cite{E895HBT,sullivan-lambda}.
Since $\lambda$ is also affected by several experiment-specific
effects (e.g. $e^-$ rejection efficiency), we do not focus on 
the details of its excitation function.

The radius parameter $R_{s}$ correlates most directly with source geometry~\cite{hbt-review}.
After an initial decrease at AGS energies, attributed to increasing space-momentum
correlations~\cite{E895HBT}, $R_{s}$ 
appears to rise slightly with collision energy, reflecting a larger freeze-out volume with
increasing pion multiplicity.
Similarly, $R_{l}$ does not exhibit a large
increase with collision energy, after the initial increase between AGS and SPS energies.

The parameter $R_{o}$ encodes both geometry and timescale information; 
in the absence of flow and opacity effects, the emission
timescale (duration of freeze-out) is given by 
$\tau = \sqrt{R_{o}^2-R_{s}^2}/\beta_T$, where $\beta_T$ is the average transverse velocity of a pion pair.
Both $R_o$ and $\sqrt{R_o^2-R_s^2}$ are relatively
independent of energy; we see no evidence for a large increase in emission
timescale (e.g. due to QGP formation~\cite{rischke-lifetime}).

In hydrodynamic models~\cite{rischke-lifetime,rischke-gyulassy-lifetime},
the ratio $R_{o}/R_{s}$ is a sensitive probe of the Equation of State (EOS).
In a purely hadronic (non-QGP) scenario, $R_{o}/R_{s}\approx 1.0-1.2$, while formation of QGP is predicted
to produce $R_{o}/R_{s}\approx 1.5-10$.  This ratio, plotted in the bottom panel of
Fig.~\ref{fig:ExcitationFctn} for low-$p_T$ pions, does not show a significant rise at RHIC.  Furthermore, at RHIC
$R_{o}/R_{s}$ is observed to decrease (from 1.07 to 0.89) with $p_T$, in contrast to recent
transport calculations~\cite{soff-bass} which include effects of hadronic rescattering after a 
hydrodynamic stage with a QGP phase.
We note, however, that the long emission duration QGP signature is expected~\cite{rischke-gyulassy-lifetime} 
to vanish again at energies higher than its onset, due to rapid expansion of the system.
Because of the large energy difference between SPS and
     RHIC, without intervening measurements, we cannot rule
     out an increase in the emission timescale at some lower,
     intermediate energy.

    In conclusion, STAR has measured two-pion correlation functions in Au+Au collisions at $\sqrt{s_{NN}}$ = 130~GeV.
Transverse HBT radii grow with event multiplicity, reflecting the evolution of source geometry with centrality.
In contrast to studies at lower energies, $R_l$ also increases steadily over a large range of multiplicities; more study is required
to understand the origin of this effect.
The $m_T$-dependence of the HBT radii is stronger than at lower energies, suggesting emission from a more rapidly expanding source at RHIC,
and in particular the much stronger decrease in $R_o$ at high $m_T$ may indicate the onset of opaqueness in the dense system.
The pion interferometry excitation function for the heaviest ions now spans nearly two decades in $\sqrt{s_{NN}}$.
No sudden jumps in HBT radii are observed, but lower energy RHIC measurements 
are needed to complete the search for a predicted increase in emission 
timescale related to the possible onset of QGP formation.

We wish to thank the RHIC Operations Group at Brookhaven National 
Laboratory for their tremendous support and for providing collisions 
for the experiment. This work was supported by the Division of Nuclear 
Physics and the Division of High Energy Physics of the Office of Science of 
the U.S.Department of Energy, the United States National Science Foundation,
the Bundesministerium fuer Bildung und Forschung of Germany,
the Institut National de la Physique Nucleaire et de la Physique 
des Particules of France, the United Kingdom Engineering and Physical 
Sciences Research Council, Fundacao de Amparo a Pesquisa do Estado de Sao 
Paulo, Brazil, and the Russian Ministry of Science and Technology.

\noindent

\end{multicols}

\end{document}